# Ultra-Low Power Crypto-Engine Based on Simon 32/64 for Energy- and Area-Constrained Integrated Systems

Sachin Taneja, Massimo Alioto

National University of Singapore

## Abstract

This paper proposes an ultra-low power crypto-engine achieving sub-pJ/bit energy and sub-1K$\mu m^2$ in 40nm CMOS, based on the Simon cryptographic algorithm. Energy and area efficiency are pursued via microarchitectural exploration, ultra-low voltage operation with high resiliency via latch-based pipelines, and power reduction techniques via multi-bit sequential elements. Overall, the comparison with the state of the art shows best-in-class energy efficiency and area. This makes it well suited for ubiquitous security in tightly-constrained platforms, e.g. RFIDs, low-end sensor nodes.

## Introduction

Energy and area efficiency are essential requirements to enable truly ubiquitous security via data encryption along the entire chain of trust, from IoT edge devices to the cloud. Existing cryptographic standards such as Advanced encryption standard (AES) [1]-[3] are currently the preferred choice for 128-bit data and key size, or higher. However, area and energy required by implementations with such data/key size are unaffordable in low-end edge devices. Also, 128-bit security is typically beyond the actual requirements of low-end devices, especially when they provide sparse and small amounts of data (e.g., real-time environmental sensors). The energy penalty due to the usage of a data wordlength beyond necessary is aggravated by the additional cost of the external FIFO buffers (~2.5Kgates and ~25-30% of area-efficient AES designs [2]) used for data word aggregation (Fig. 1).

Recently, various lightweight ciphers were proposed to relax the energy and area requirements of on-chip encryption, such as Simon [4] and PRINCE [5]. Also, Simon provides a variety of data and key sizes, instead of being fixed. Hence, in applications requiring short data wordlength, the Simon 32/64 algorithm (i.e., 32-bit data, 64-bit key) eliminates the need for FIFOs at 32-bit wordlength, and drastically reduces the FIFO size at lower wordlengths, compared to 128-bit algorithms.

In this paper, we propose a cryptographic engine based on Simon 32/64 that is a better fit for the typical data wordlength of IoT sensor nodes and embedded microcontroller, and eliminates the significant area and energy overhead of the FIFO buffers. The adoption of various techniques from energy-aware architectural selection to multi-bit pulsed latch pipelines allows operation at sub-pJ/bit energy and 1.2k-gate complexity.

## Proposed Crypto Engine Architecture

The microarchitecture plays an important role in defining the energy and the area efficiency of the cryptographic engine. Fig. 2(a) shows the area and energy (post-place&route) of the bit-serial (1 bit per clock) and the bit-parallel (32 bits, i.e. 1 round per cycle) design of the Simon engine. Post layout simulations show that the adoption of the bit-parallel microarchitecture brings an energy efficiency improvement by 16X and area savings of 9%, compared to the bit-serial microarchitecture. Energy and area savings are justified by the fact that the area and energy overhead of the parallel datapath due to the combinational logic is much lower than the cost of bit serialization via additional flip-flops and control logic.

As shown in Fig. 2(b), the Simon bit-parallel microarchitecture tends to be dominated by flip-flops, in terms of both area (50%) and energy (65%). The clocking energy contribution was reduced through the adoption of the pulsed latch as sequencing element, as shown in Fig. 3(a). Indeed, a pulsed latch occupies 25% less area, 40% lower clock pin energy, and 20% lower energy per cycle than a flip-flop. In view of the dominance of sequencing elements in the Simon engine, the lower area of pulsed latches permits to shrink the overall area and hence reduce the switched capacitance in both the datapath and the clock tree. The adoption of pulsed latches reduces energy and also enables time borrowing providing resiliency against process/voltage/temperature variations [6].

Additional energy and area savings were achieved by introducing multi-bit pulsed latches, i.e. sharing the same clock drivers across multiple pulsed latches as shown in Fig. 3(b). The internal clock buffer was sized (2X) to balance the internal clock slope (determining hold time and energy), and the cell area. This assures reduction in clocking energy by 40% as compared to 1-b latch-based design (see Fig. 4). The resulting 8-bit pulsed latch design was created to enrich a commercial standard cell library, and to be integrated with an existing digital design flow.

## Post Layout Simulation Results

The post-layout simulation results in Fig. 4 show the area and energy benefits (12-14%) with adoption of multi-bit pulsed latches as compared to conventional flip-flops. Fig. 5 describes the final bit-parallel microarchitecture (datapath and key expansion) employing the Simon 32/64 engine. The proposed Simon 32/64 engine is compared in table I with the baseline Simon 32/64 with bit-serial architecture [4] and flip-flop as sequencing element. Both the proposed and the baseline engine were designed using the same design flow in 40nm CMOS.

The proposed bit-parallel microarchitecture occupies 690$\mu m^2$ (Fig. 6). The maximum throughput at 0.9V, 25°C and typical corner is 443Mbps without time borrowing (which offers an extra 10% cycle reduction). The power consumption is 434$\mu W$ at 443MHz, leading to an energy efficiency of 1.02Tbps/W. No timing violations occur under a duty cycle (externally generated) of 10-24%. The minimum energy point lies at 225mV, leading to an energy per bit of 0.104pJ/b.

The baseline bit-serial microarchitecture occupies 861$\mu m^2$ (20% more than the proposed), supports maximum throughput of 22.7Mbps at 0.9V, 25°C and typical corner. Power is 424$\mu W$ at 364MHz, leading to an energy efficiency of 53Gbps/W. The proposed cryptographic engine is compared with the state of the art [2]-[9] in table II. It has the lowest area and energy consumption, making it suited for tightly resource-constrained platforms.

## Acknowledgements

Authors thank Singapore MOE grant MOE2016-T2-1-150.

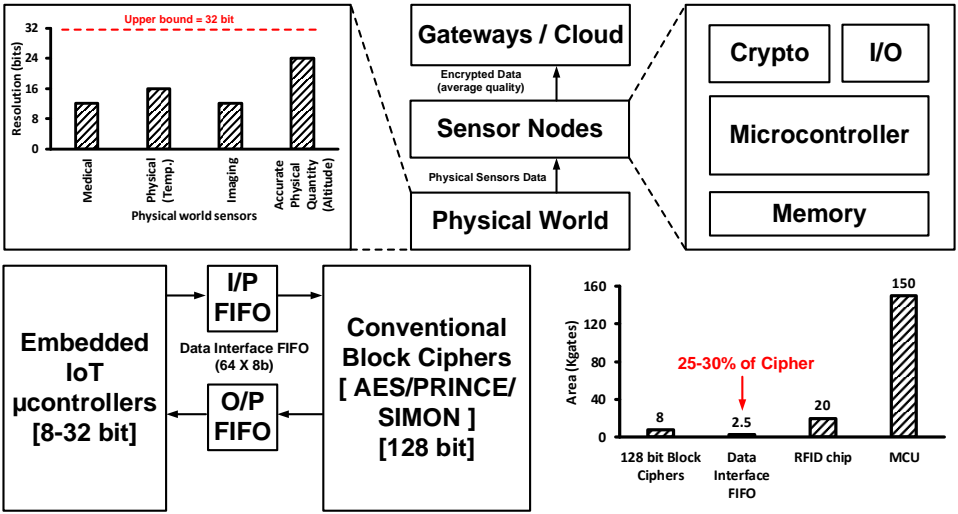
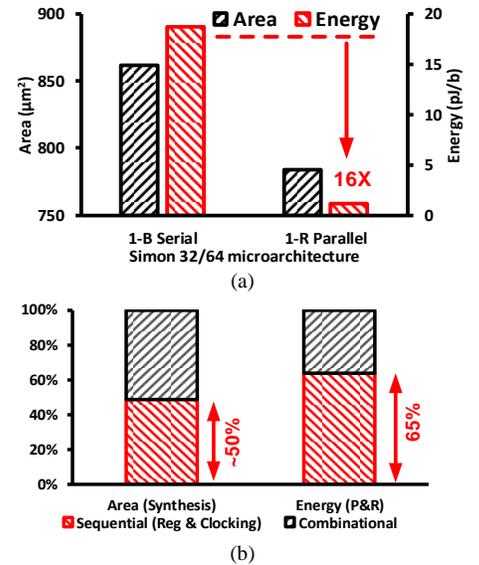

Fig. 1. Symmetric key cryptography requirements for constrained platforms (e.g., RFIDs, IoT nodes).

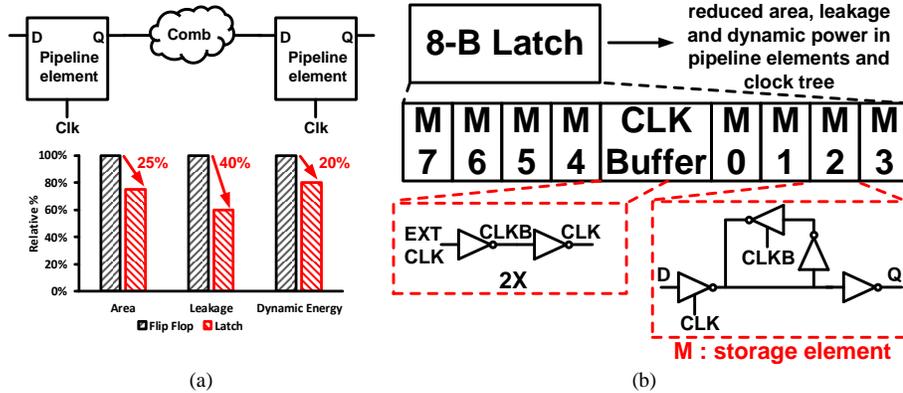
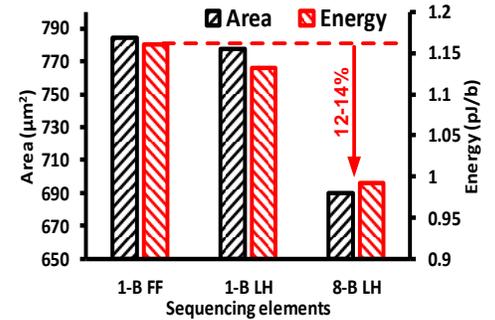

Fig. 2. (a) Comparison of bit-serial and bit-parallel microarchitecture, (b) Sequential and combinational breakdown.

Fig. 3. (a) Pulsed latches advantages in area, leakage and energy with time borrowing over flip-flops as sequencing elements, (b) Multi-bit latch design and advantages.

Fig. 4. Area and energy improvement by adoption of multi-bit latch-based pipelines.

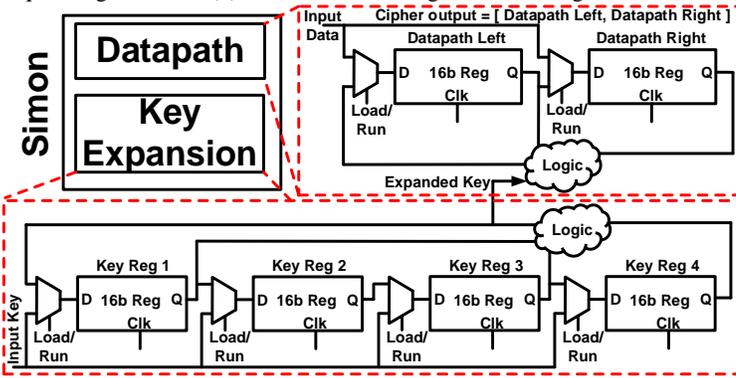

Fig. 5. Proposed bit-parallel microarchitecture of Simon 32/64.

Table I. Summary of proposed and baseline Simon architecture.

| Architecture | Proposed Simon | Baseline Simon |
|---|---|---|
| Pipeline | 8-bit latch | 1-bit flip flop |
| Clocking | Pulsed (duty cycled) | Edge |
| Microarchitecture | Bit-parallel | Bit-serial |
| Core area (µm²) | 690 | 861 |
| Maximum throughput (Mbps) | 443 | 22.7 |
| Power @ fmax (µW) | 434 | 424 |
| Energy/bit (pJ/b) | 0.99 | 18.69 |

Table II. Comparison of proposed work with state of the art cryptographic engine (at nominal voltage).

| | This work (proposed) | JIOT' 2018 [9] | ISCAS' 2018 [8] | VLSIC' 2017 [7] | VLSIC' 2016 [3] | TVLSI' 2015 [2] |
|---|---|---|---|---|---|---|
| Algorithm | SIMON | SIMON [a] | SIMON [b] | PRINCE | AES | AES |
| Datapath size (bits) | 32 | 128 | 32 | 128 | 128 | 128 |
| Key size (bits) | 64 | 128 | 64 | 128 | 128 | 128 |
| Technology (nm) | 40 | 15 | 65 | 28 | 40 | 65 |
| Area / F² (X 10⁶) | 0.43 | 3.23 | - | 9.44 | 2.68 | 1.89 |
| Nominal voltage (V) | 0.9 | 0.8 | 1.2 | 1 | 0.9 | 0.7 |
| Energy/bit (pJ/b) | 0.99 | 0.482 | 1.06 | 0.39 | 8.85 | 3 |
| Maximum throughput (Mbps) | 443 | 31,800 | 0.616 | 25000 | 432 | 100 |
| Energy efficiency/Area (Gbps/W.µm²) | 1.464 | 2.854 | 2.247 | 0.346 | 0.026 | 0.042 |
| Throughput efficiency (Mbps/µm²) | 0.642 | 43.741 | 0.001 | 3.377 | 0.101 | 0.013 |

a. Simon 1 round per clock design, only synthesis (w/o Place and Route) results
b. PAL logic style, 41% efficiency (energy calculation) omitting area overhead for clock, and ignoring layout parasitics
F: Minimum feature size of technology

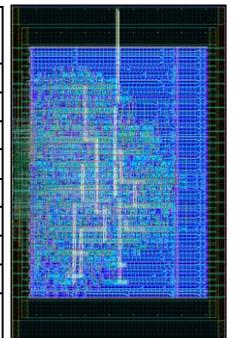

Fig. 6. Proposed cryptographic engine layout in 40nm.